\documentclass[aps,prb,reprint,a4paper,floatfix,superscriptaddress,amsmath,amssymb,amsfonts]{revtex4-1}
\usepackage{mathptmx}
\usepackage[scaled=0.9]{helvet}
\usepackage[utf8]{inputenc}
\usepackage{graphicx}
\usepackage{textcomp}
\usepackage{upgreek} 
\usepackage[dvipsnames]{xcolor}
\usepackage{soul} 
\usepackage{xspace}
\usepackage{array}
\usepackage{makecell}
\usepackage[export]{adjustbox}

\usepackage[pdftex]{hyperref}
\hypersetup{
	pdftitle={Influence of strain relaxation in axial In$_x$Ga$_{1-x}$N/GaN nanowire heterostructures on their electronic properties},
	colorlinks,
	citecolor=blue
	}

\usepackage[
,textwidth=17.5cm
,textheight=23.5cm
,verbose
,pdftex
]{geometry}

\setcitestyle{numbers,square}
\begin{document}
\makeatletter
\renewcommand\@biblabel[1]{\mbox{$\left[ \right.$}#1\mbox{$\left. \right]$}}
\makeatother

\title{Influence of strain relaxation in axial In$_x$Ga$_{1-x}$N/GaN nanowire heterostructures on their electronic properties}

\author{Oliver Marquardt}
\email{marquardt@pdi-berlin.de}
\author{Thilo Krause}
\author{Vladimir Kaganer}
\affiliation{Paul-Drude-Institut f\"ur Festk\"orperelektronik, Hausvogteiplatz 5-7, D-10117 Berlin, Germany}
\author{Javier Mart\'in-S\'anchez}
\affiliation{ Institute of Semiconductor and Solid State Physics, Johannes Kepler University Linz, Altenbergerstr.\ 69, A-4040 Linz, Austria }
\author{Michael Hanke}
\affiliation{Paul-Drude-Institut f\"ur Festk\"orperelektronik, Hausvogteiplatz 5-7, D-10117 Berlin, Germany}
\author{Oliver Brandt}
\affiliation{Paul-Drude-Institut f\"ur Festk\"orperelektronik, Hausvogteiplatz 5-7, D-10117 Berlin, Germany}

\date{\today}

\begin{abstract}
We present a systematic study of the influence of elastic strain relaxation on the built-in electrostatic potentials and the electronic properties of axial In$_x$Ga$_{1-x}$N/GaN nanowire heterostructures. We employ and evaluate analytical and numerical approaches to compute strain and polarization potentials. These two ingredients then enter an eight-band $\mathbf{k}\cdot\mathbf{p}$ model to compute electron and hole ground states and energies.
Our analysis reveals that for a sufficiently large ratio between the thickness of the In$_x$Ga$_{1-x}$N disk and the diameter of the nanowire, the elastic relaxation leads to a significant reduction of the built-in electrostatic potential in comparison to a planar system of similar layer thickness and In content.
However, a complete elimination of the built-in potential cannot be achieved in axial nanowire heterostructures. Nevertheless, the reduction of the built-in electrostatic potential leads to a significant modification of the electron and hole energies.  Our findings indicate that the range of accessible ground state transition energies in an axial In$_x$Ga$_{1-x}$N/GaN nanowire heterostructure is limited due to the reduced influence of polarization potentials for thicker disks.
Additionally, we find that strain and polarization potentials induce complex confinement features of electrons and holes, which depend on the In content, shape, and dimensions of the heterostructure.
\end{abstract}


\maketitle

The ternary alloy In$_x$Ga$_{1-x}$N constitutes the semiconductor material of choice for the development of red-green-blue light emitting diodes (LEDs) for display technology, since its emission wavelength can in principle be tuned from the near-infrared to the ultraviolet via the In content $x$~\cite{NaFa97, DaGr00, RiWi08}. However, due to the large lattice mismatch of about 10\% between InN and GaN and the tendency for phase separation, it is difficult to produce In$_x$Ga$_{1-x}$N films on GaN with the In content required for red emission while retaining a sufficiently high crystal quality \cite{SiDo97}. Additionally, the strong polarization potentials occuring in planar In$_x$Ga$_{1-x}$N/GaN heterostructures induce a spatial separation of electrons and holes and correspondingly low recombination rates.

A possible solution to overcome the limitations of planar heterostructures is the growth of GaN nanowires (NWs) with axial In$_x$Ga$_{1-x}$N insertions~\cite{KiKa04,KiCh04,LiWa12}. In contrast to planar structures, the large surface-to-volume ratio facilitates elastic relaxation of the lattice mismatched axial insertions~\cite{BjOh02,ErGr05,HaEi07,KaBe12}, thus making the incorporation of larger In contents in the insertions possible without inducing plastic relaxation.
Along with the elastic relaxation, a reduction of the built-in piezoelectric potential and thus the quantum confined Stark effect~\cite{MiCh84} (QCSE) is expected. In fact, some researchers even reported evidence for a vanishing QCSE~\cite{ArTs10,LiLu10,NgZh11,BaKa09}. The influence of elastic relaxation of axial In$_{x}$Ga$_{1-x}$N disks on the piezoelectric potential in GaN NWs has only recently been subject of a systematic study, and it was shown that a total elimination of piezoelectric fields cannot be achieved in axial NW heterostructures~\cite{KaMa16}. However, the question of how this elastic relaxation influences carrier confinement and transition energies has not been addressed so far.

In the following, we will shed light on the impact of elastic relaxation in axial In$_x$Ga$_{1-x}$N/GaN NW heterostructures on their piezoelectric and electronic properties. We focus particularly on thin nanowires, where the ratio between the thickness of the In$_x$Ga$_{1-x}$N insertion and the NW radius approaches unity, such that significant elastic relaxation is expected~\cite{KrHa16}. We report a strongly reduced magnitude of the polarization potential in the NW heterostructure in comparison with a planar structure of similar thickness and In content.  Our findings indicate that the range of accessible transition energies in axial In$_x$Ga$_{1-x}$N/GaN NW heterostructures is limited in comparison to planar layers due to the reduced influence of polarization potentials. in comparison to planar layers


\begin{table}[tbp]
\caption{Overview of the three model cases A, C, and H.}
\begin{ruledtabular}
\begin{tabular}{ccc}
 \multicolumn{1}{c}{Computational approach} & \multicolumn{2}{c}{Nanowire geometry} \tabularnewline
~ & Cylindrical & Hexagonal\tabularnewline
 \hline
Analytical	& A & ---\tabularnewline
Numerical	& C & H\tabularnewline
\end{tabular}
\label{tab:model}
\end{ruledtabular}
\end{table}

Our model system is a GaN NW with a diameter of 15~nm containing an In$_{x}$Ga$_{1-x}$N disk of thickness $t$.  Such \textit{ultrathin} GaN NWs have been recently fabricated by either direct growth~\cite{ChTs12} or post-growth thermal decomposition~\cite{ZeCo16}. We compute the strain state of this NW based on continuum elasticity theory~\cite{LaLi86}.
Presuming the  NWs to exhibit a cylindrical shape and to have the same elastic, piezoelectric and dielectric constants throughout the whole axial NW heterostructure, we compute strain and polarization potentials analytically as outlined in Refs.~\cite{KaBe12} and \cite{KaMa16}.  With these simplifications, we achieve exact solutions of both the elastic problem and the Poisson equation.  In the following, we refer to this approach as 'case A' (see Tab.~\ref{tab:model}).
A more realistic description of the NWs can be achieved using the numerical finite element method (FEM). We first start with FEM-based simulations of a NW that maintains the cylindrical shape, but allows us to take into account a spatial variation of the elastic, piezoelectric, and dielectric constants. We refer to this approach as 'case C'.
Finally, a more realistic, hexagonal shape of the NW is considered within the FEM approach ('case H'). To compute strain and polarization potentials for the numerical models in cases C and H, we employ the commercial FEM solver MSC Marc\textsuperscript{\textregistered}~\footnote{{Note that the off-diagonal strains reported in Ref.~\cite{KrHa16} have to be divided by 2 before computing their contribution to the polarization potential due to the internal notation of the software.}}. In the present work, we provide a comparison of the three approaches with a focus on the polarization potentials and the electronic properties.

Strain and piezoelectric potentials obtained from the above models enter the calculation of the ground-state electron and hole single particle wave functions and binding energies using an eight-band $\mathbf{k}\cdot\mathbf{p}$ model for wurtzite crystals~\cite{chch96}.  These simulations were performed using the S/PHI/nX library~\cite{BoFr11,MaBo14}.  A comparison of the ground state recombination energies was performed with planar In$_{x}$Ga$_{1-x}$N/GaN heterostructures for different disk thicknesses. For the sake of comparability, strain and polarization of the planar layer were transferred to a structure with the same cross section as the NW, such that small energy contributions from in-plane quantization are consistent in both the NW and the planar system.
The material parameters for our simulations were taken from Ref.~\cite{WiSc06}. However, following a previous study~\cite{ScCa11}, we use a negative value for $e_{15}$ = $e_{31}$.
Note that the analytic approach yields the same results as the numerical FEM approach if elastic, piezoelectric and dielectric constants are assumed to be the ones of GaN throughout the whole NW.


\begin{figure}
\includegraphics[width=\columnwidth]{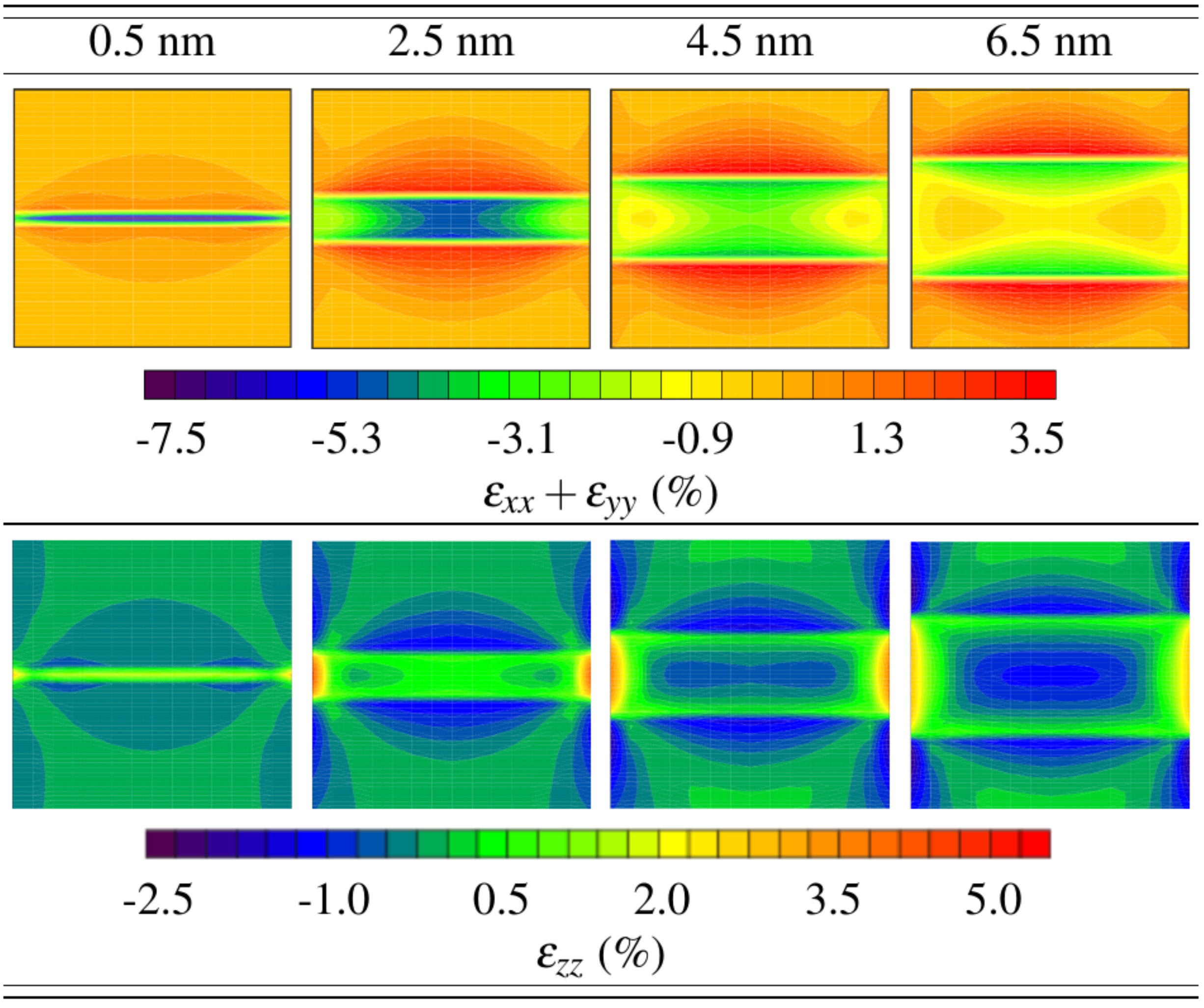}
\caption{(Color online) Cross section of in-plane (top) and out-of-plane (bottom) strain components for different thicknesses of an In$_{0.4}$Ga$_{0.6}$N disk embedded in a GaN NW (case H).\label{fig:strain}}
\end{figure}

To illustrate the elastic relaxation of an axial In$_x$Ga$_{1-x}$N/GaN NW heterostructure, we show in Fig.~\ref{fig:strain} the diagonal components of the elastic strain tensor, $\varepsilon_{zz}$ and $\varepsilon_{xx}+\varepsilon_{yy}$, for the model NW with a hexagonal cross section (case H) for an In content of 40\%. A significant relaxation is seen for thicknesses of 4.5 and 6.5~nm.  For thin disks, lateral relaxation occurs only at the side facets of the NW so that the central area of the In$_x$Ga$_{1-x}$N disk is subject to elastic strain that is very similar to the one of a planar layer. In all cases, however, the lattice mismatch between In$_{x}$Ga$_{1-x}$N and GaN induces significant strains at the interfaces which will modify the polarization potentials and thus the electronic properties of the NW. The strain profiles obtained from cases A and C are very similar to those of case H except for slightly larger local strains occuring in the corners of the hexagonal NWs in case H as compared to the strains at the side facets.


\begin{figure}
\includegraphics[width=\columnwidth]{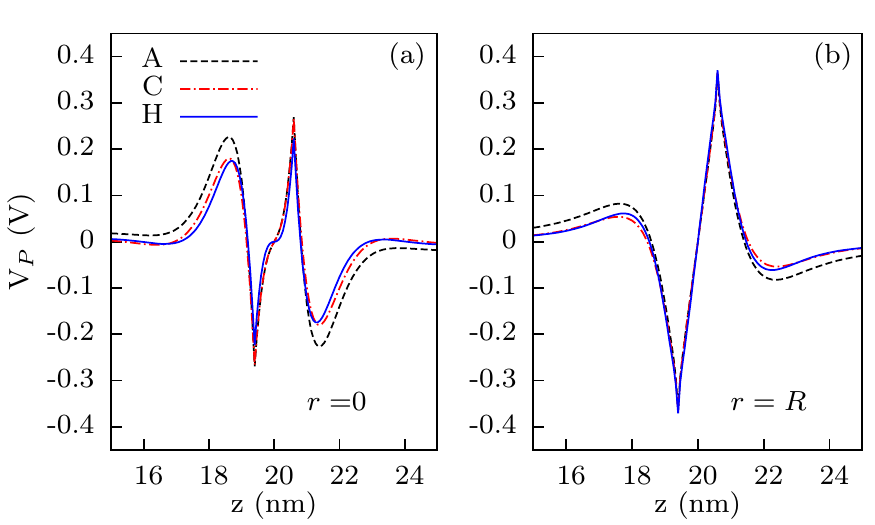}
\caption{(Color online) Line-scans of the polarization potential in an In$_{0.4}$Ga$_{0.6}$N/GaN NW heterostructure with a diameter of 15~nm and a thickness of the In$_{0.4}$Ga$_{0.6}$N disk of 4.5~nm computed for the cases A, C, and H along the central axis of the NW at $r$ = 0 (a) and at the side facet (cases A and C) or edge (case H) of the NW at $r$ = $R$ (b).}\label{fig:piezolines}
\end{figure}
We now compare the polarization potentials obtained for cases A, C, and H. Linescans of the polarization potential along the central axis as well as along a parallel line at the surface ($r$ = $R$) are shown in Fig.~\ref{fig:piezolines} for a 4.5~nm thick In$_{0.4}$Ga$_{0.6}$N disk embedded in a GaN NW of 15~nm in diameter. We observe that the three models yield close results, with deviations occuring mainly outside the In$_x$Ga$_{1-x}$N disk near the interfaces for case A. These deviations are a result of the fact that for this approach, the elastic, piezoelectric, and dielectric constants throughout the whole NW are assumed to be identical to those of GaN.
In the present case, we found the elastic constants to be largely responsible for the differences in the piezoelectric potential outside the active layer in Fig.~\ref{fig:piezolines}.
Deviations are visible particularly for larger In contents, where the assumption of spatially constant material parameters (i.e., those of GaN) becomes increasingly inaccurate. However, the polarization potentials for cases H and C with spatially dependent material parameters are in very good agreement. Hence, the shape of the NW appears to play only a minor role.
The extrema of the polarization potential along an axis parallel to the NW growth direction are slightly larger near the corners as compared to the ones near the center of a side facet (not shown in Fig.~\ref{fig:piezolines}).

\begin{figure}
\includegraphics[width=\columnwidth]{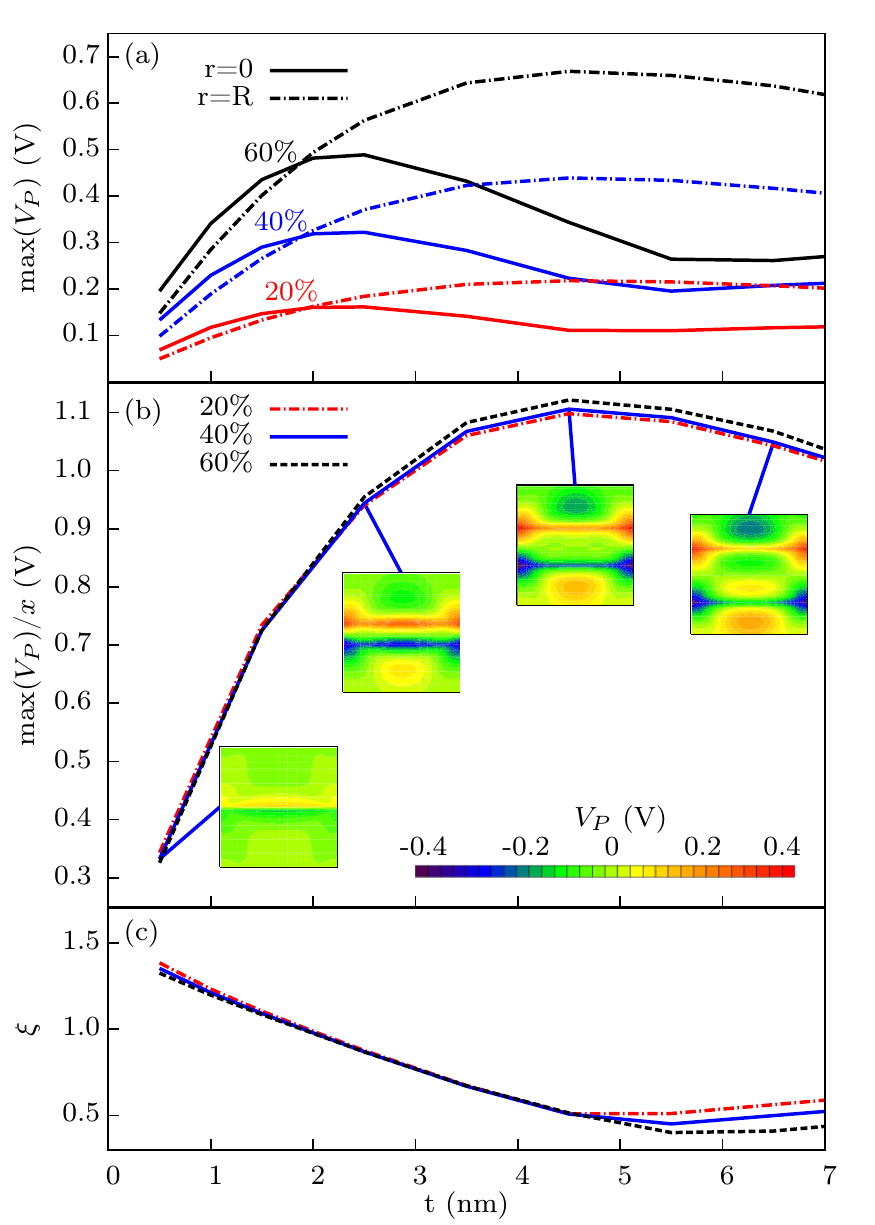}
\caption{(Color online) (a) Maximum of the polarization potential along the axis of the NW (solid) and along a parallel line at the side facet ($r$ = $R$, dash-dotted) as a function of the disk thickness for case H and In contents of 20 (red), 40 (blue), and 60\% (black). (b) Overall maximum of the polarization potential divided by the In content as a function of the disk thickness for case H for In contents of 20, 40, and 60\%. Insets show a side-view of the polarization potential for selected thicknesses for an In content of 40\%. (c) Ratio between the local maxima of the polarization potential at the center and at the side facets of the NW as a function of the disk thickness.}\label{fig:piezo}
\end{figure}
Fig.~\ref{fig:piezo} (a) shows the maximum of the polarization potential along the axis of the NW as well as along a parallel line at the side facet for different In contents as a function of the disk thickness. The maxima along the NW axis are largest for all In contents at a thickness of 2.5~nm and then decreases to a minimum value at 4.5 ($x = $ 20\%), 5.5 ($x = $ 40\%), and 6.5~nm ($x = $ 60\%). After this point, the maximum of the polarization potential along the NW axis increases again due to the spontaneous polarization. At the side facet, the maximum of the potential has its largest value for disk thicknesses of 4.5~nm.

The polarization potential scales almost linearly with the In content of the In$_x$Ga$_{1-x}$N disk. The qualitative behavior of the polarization potential as a function of the disk thickness can therefore be described better after normalizing the potential with the In content.  The overall maximum of the normalized polarization potential throughout the whole NW in case H with a NW diameter of 15~nm is shown in Fig.~\ref{fig:piezo} (b) as a function of the disk thickness. The absolute extrema of the polarization potential have a maximum at a disk thickness of about~4.5~nm and a reduction is seen after this point. This behavior translates into a limitation of the range of electron-hole transition energies, as we will discuss in the following. It is also observed that the absolute maximum of the polarization potential scales with the In content and only small modifications of the curve are seen for different In contents. For case A, the curves for different In contents are completely identical as the polarization potential scales strictly linearly with the In content in this analytical approach. Apart from this fact, only small deviations are observed for the curves shown in Fig.~\ref{fig:piezo} (b) when considering cases A and C in comparison to case H. The insets in Fig.~\ref{fig:piezo} (b) show that the potential has its extrema at the central axis of the NW for thin disks and at the side facets for thick disks. This behavior is quantitatively depicted in Fig.~\ref{fig:piezo} (c), which shows the ratio $\xi=\max[V_P(0)] / \max[V_P(R)]$ between the maximum potential along the central axis and the one along a parallel axis at the side facets as a function of the disk thickness. Above a thickness of about~2.8~nm, the extrema of the potential are at the side facets rather than at the central axis of the NW. The splitting of $\xi$ for the different In contents above a thickness of 4.5~nm result from the minima of max($V_P$) along the NW axis as shown in Fig.~\ref{fig:piezo} (a), which occur at different thicknesses for the In contents considered.


\begin{figure}
\includegraphics[width=\columnwidth]{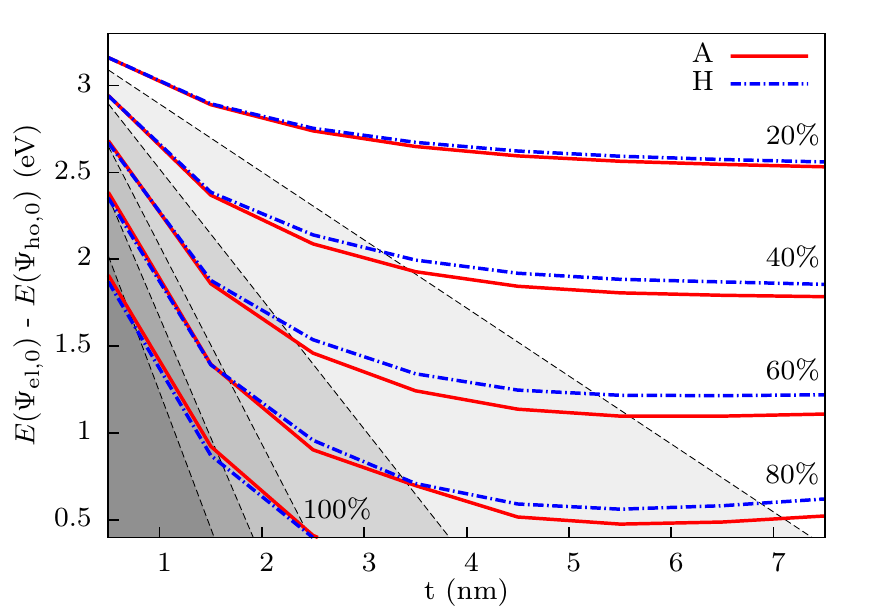}
\caption{(Color online) Electron-hole ground state transition energy in the In$_{x}$Ga$_{1-x}$N/GaN NW heterostructure under consideration for cases A (red solid) and H (blue dash-dotted) as a function of the disk thickness. The In content $x$ is indicated at each curve. For comparison, the transition energies for the limiting case of a planar heterostructure with the same In contents are indicated in shades of gray from light (20\%) to dark (100\%).}\label{fig:energies}
\end{figure}

Figure~\ref{fig:energies} shows the energy difference between the electron and hole ground states, $\Psi_\mathrm{el,0}$ and $\Psi_\mathrm{ho,0}$, for cases A and H as a function of the disk thickness in comparison with those obtained for a planar system of the same In content (shaded gray areas). The overall agreement between the models A and H is good. Case C (not shown here) deviates from case H by only a few meV. The deviation between A and H increases initially with larger In contents due to the fact that the assumption of constant material parameters throughout the whole system becomes increasingly inaccurate. For even larger In contents (80 and 100\%), however, the deviation decreases again due to the small absolute value of $E(\Psi_\mathrm{el,0}) - E(\Psi_\mathrm{ho,0})$.

Evidently, the evolution of the transition energy with increasing disk thickness $t$ differs drastically between the planar and the nanowire heterostructures. For the former, the transition energies decrease linearly with layer thickness due to the QCSE, while for the latter, the transition energies are seen to saturate for thicker disks. As a result, long wavelength emission requires a significantly higher In content than for the corresponding planar heterostructure. For example, red emission at about 1.9~eV is predicted to be achieved with a planar heterostructure containing 20\% In at a layer thickness of about 3.5~nm. A NW heterostructure emitting at this wavelength and at the same layer thickness would require an In content of about 45\%. Alternatively, one could employ a 5.5~nm thick disk with an In content of 40\% or a 1.5~nm thick disk with an In content of 60\%.
\begin{figure}[h]
\includegraphics[width=\columnwidth]{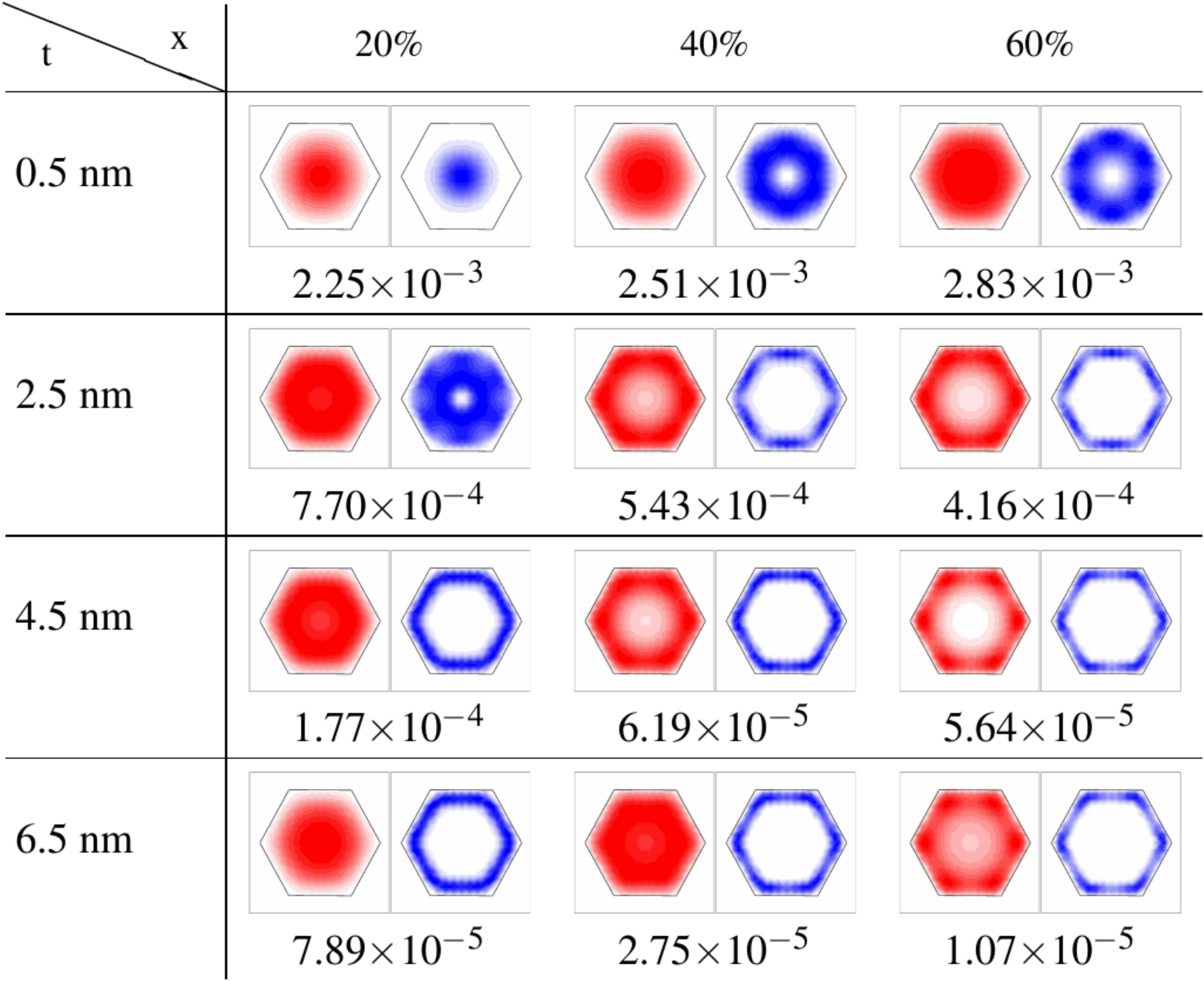}
\caption{(Color online) Electron (red) and hole (blue) ground state charge densities for different In contents and disk thicknesses in top view. The number below the charge densities shows the respective overlap $\mathcal{O}$.\label{fig:charges}}
\end{figure}

Since we have assumed that the NW has a diameter of only 15~nm, the strain induced by the In$_x$Ga$_{1-x}$N disk should be accomodated elastically for the entire parameter range depicted in Fig.~\ref{fig:energies}~\cite{Glas06}. This elastic strain relaxation strongly reduces the polarization potential in the nanowire heterostructure as compared to the equivalent planar heterostructure [cf.\ Fig.~\ref{fig:piezo} (b)], but at the same time also reduces the redshift induced by the QCSE, requiring in turn significantly larger amounts of In for obtaining the same emission wavelength. The crucial question at this point is whether or not the reduced piezoelectric potential increases the electron-hole overlap and thus constitutes an actual advantage. Let us consider the example given above of structures emitting at 1.9~eV. For the planar heterostructure, the electron-hole overlap $\mathcal{O}$ as defined in~\cite{MaHa13} is extremely small (1.17$\times$10$^{-9}$) due to the strong polarization potential. In contrast, we obtain values of 4.06$\times$10$^{-5}$ and 1.56$\times$10$^{-3}$ for the nanowire heterostructures with 40\% In content and 5.5~nm thickness and 60\% In content and 1.5~nm thickness, respectively. For the thin disk with 60\% In content, the large overlap results from the negligible spatial separation of electron and hole along the growth direction. However, even for an In content of only 40\%, the electron-hole overlap is significantly larger than for a planar layer. 

In general, however, we cannot predict simple monotonic trends for the electron-hole overlap as a function of disk thickness and In content, but need to examine each nanowire heterostructure individually. To illustrate this fact, we show the electron and hole ground state charge densities together with the respective overlap $\mathcal{O}$ for a few selected configurations in Fig.~\ref{fig:charges}.
The electron is typically confined in the center of the NW for both thin disks and low In contents, but tends to be confined in the corner of the NW for thicker disks and larger In contents. For the hole, a confinement in the center of the NW is observed only for an In content of 20\%. Otherwise, the hole is confined at the corner (e.g., for disk thicknesses of 4.5 and 6.5~nm for an In content of 60\%) or the side facets (for disk thicknesses of 2.5~nm for $x=40$\% and 60\% and for disk thicknesses of 4.5 and 6.5 nm for an In content of 40\%), as also reported in Ref.~\onlinecite{BoVe10}. Note that in all cases the electron is confined at the top facet of the active In$_x$Ga$_{1-x}$N disk, whereas the hole resides at its bottom, due to the drop of the polarization potential along the NW axis seen in Fig.~\ref{fig:piezolines}. This vertical separation leads to a reduction of the overlap for larger disk thicknesses, as seen in Fig.~\ref{fig:charges}.

To retain an electron-hole overlap higher than that for a planar heterostructure with comparable transition energy, it is imperative to chose a configuration for which no radial separation of electrons and holes occurs. This effect, however, evidently depends on the structural parameters of the nanowire heterostructure in a highly nontrivial manner. In the present study, we have considered the In$_x$Ga$_{1-x}$N insertion to be represented by a disk, but it is known from experiment that the insertion may assume rather complex shapes that have been found to affect the spatial distribution of the electron and hole charge densities as well~\cite{MaGe15}. To predict the transition energy and the electron-hole overlap of a specific nanowire heterostructure would require a complete three-dimensional reconstruction of the insertions’s size, shape and composition on a nm scale. Electron and atom-probe tomography are experimental techniques offering the possibility of such a reconstruction.        

\acknowledgements
The authors would like to thank Lutz Geelhaar for valuable suggestions and Ryan B. Lewis for a critical reading of the manuscript.

\bibliography{paper}
\end{document}